\documentclass[journal]{IEEEtran}

\usepackage[final]{graphicx}
\usepackage{amsmath,amssymb}
\usepackage{bm}
\usepackage[hidelinks]{hyperref}
\usepackage{physics}
\usepackage{adjustbox}
\usepackage{caption}
\usepackage{subcaption}
\usepackage{float}
\usepackage{cite}
\usepackage{xcolor}

\title{A neural-astrocyte architecture implements a hybrid automaton for evidence accumulation}

\author{Giacomo Vedovati, Ilya E. Monosov, Thomas J. Papouin, and ShiNung Ching
\thanks{G. Vedovati and S. Ching are with the Department of Electrical and Systems Engineering, Washington University in St. Louis, St. Louis, MO 63130, USA (e-mail: g.vedovati@wustl.edu; shinung@wustl.edu).}
\thanks{I. E. Monosov is with the Department of Neuroscience, Johns Hopkins University, Baltimore, MD 21218 USA (e-mail: imonosov@jhu.edu).}
\thanks{T. J. Papouin is with the Department of Neuroscience, Washington University in St. Louis, St. Louis, MO 63130, USA (e-mail: thomas.papouin@wustl.edu).}}

\begin{document}

\maketitle

\begin{abstract}
Astrocytes are non-neuronal glial cells that are receiving widespread attention due to their emerging role in neural computation. In this paper, we propose and study dynamical mechanisms by which astrocytes may augment the ability of neural networks to infer context in reinforcement learning (RL) settings. 
We construct a biologically inspired, two-level dynamical neural-astrocyte network with distinct spatial and temporal organization. We train this model on a hierarchical multi-context task that requires the agent to infer changes in latent task rules based on derived rewards. We find that in this setting, astrocytes enable evidence accumulation of changes in context and subsequent context-specific modulation of neural dynamics. We show that these functions are implemented via two dynamical mechanisms: (i) reward-induced bifurcations that relocate an asymptotically stable attractor into different, context-specific regions of state space, and (ii) the relative shallowness of these attractors, mediated by the entropy of the environment, giving rise to behavioral stickiness.
Together, these mechanisms amount to a hybrid automaton, in which uncertainty accumulates until, eventually, the neural dynamics are switched to a new context.
This model provides a neuro-dynamic schema, compatible with neural-astrocyte biology and prior empirical observations, for how astrocytes may integrate information from the periphery and drive contextual changes in neural circuits.

\end{abstract}

\begin{IEEEkeywords}
Neural-astrocyte networks, evidence accumulation, contextual discovery, recurrent neural networks.
\end{IEEEkeywords}

\section{Introduction}

\subsection{Neural-Astrocyte interactions and computation}
Brain-inspired network architectures have historically prioritized neurons as the primary unit of computation and information processing. However, current research is beginning to pivot toward a more holistic view of brain-inspired computing, and especially the role of non-neuronal cell types such as astrocytes.

The proportion of astrocytes in the brain is commensurate with that of neurons, and their spatiotemporal properties offer the possibility of functional complementarity. Neuronal circuits operate on the millisecond timescale, whereas astrocytes function on the second to minute timescale \cite{dong2026astrocyte}. Based on this timescale separation, a growing body of empirical work indicates a neurocomputational schema wherein fast, task‑focused neuronal circuits are embedded within slower, spatially extended modulatory networks mediated by astrocytes \cite{gong2024astrocytes,ali2025dialogue,juliaPai}. 
Importantly, astrocytes possess the mechanistic repertoire to act as integrative modulators of neural function: they sense local activity (via neurotransmitter and neuromodulator receptors), exhibit long‑timescale Ca$^{2+}$ dynamics \cite{arizono2020structural, goenaga2023calcium, khakh2025astrocyte}, release transmitters (e.g., glutamate, ATP) that modulate synaptic function \cite{de2023specialized, sahlender2014we}, participate in potassium buffering that alters excitability \cite{santello2019astrocyte,semyanov2020making,wang2012bergmann}, regulate synaptic efficacy and ionic milieu \cite{lee2021astrocytes, wang2012astrocytes}, and tile microcircuits to exert spatially specific control over neuronal dynamics \cite{bushong2002protoplasmic, paukert2014norepinephrine}. Related to this latter point, astrocytes also form gap‑junction–coupled networks that span many neurons and integrate signals over broad spatial domains and slow timescales, providing a substrate for spatially extended contextual modulation \cite{houades2008gap}.

These results support the view of astrocytes as modulatory multiplexers within the brain, capable of gating and biasing the fast neuronal networks that perform task computations \cite{murphy2022conceptual}, in response to both external (e.g., sensory signals, task instructions, environmental regularities) and internal  (e.g., behavioral outcomes, internal state, neuromodulatory tone) factors (Fig. \ref{fig:contextualGuidance}).

\subsection{Astrocytic contextual guidance and reinforcement learning}

\begin{figure}[!t]
\centering
\includegraphics[width=0.85\columnwidth]{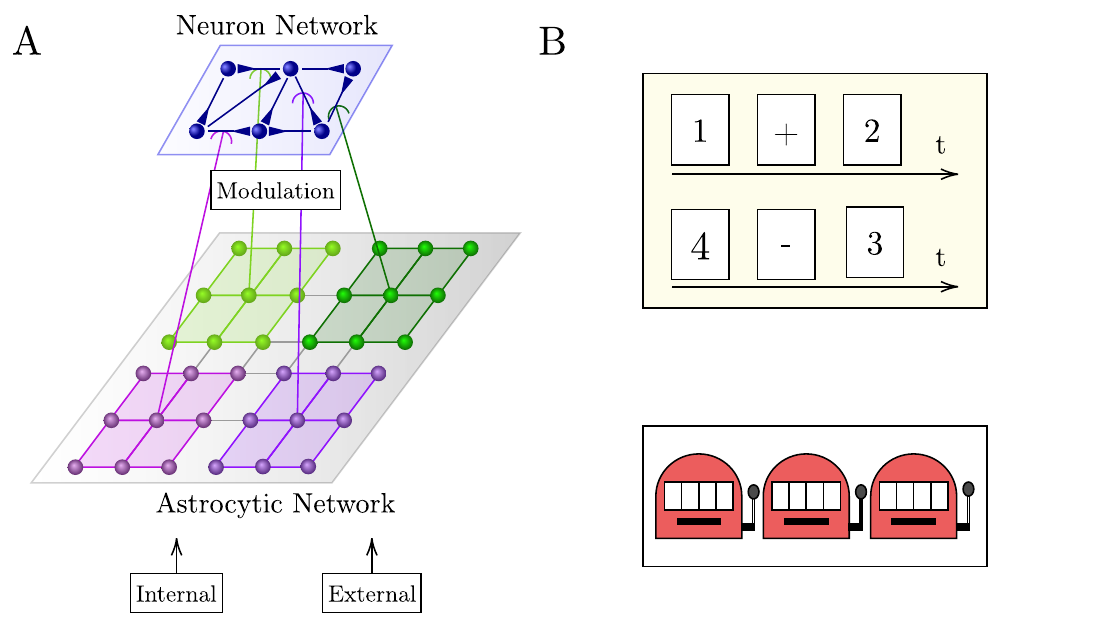}
\caption{Hierarchical neuro-astrocytic network and contextual task regimes.
A. The two-tier network architecture. The high-level Astrocytic Network (A-RNN, bottom) integrates both external (e.g., sensory signals) and internal (e.g., reward/futility feedback) factors to modulate effective synaptic weights ($\mathcal{W}_{\text{eff}}$) within the fast Neuron Network (N-RNN, top). 
B. Task regime paradigms: in explicit tasks, rules are directly driven by observable external stimuli (e.g., arithmetic); in the implicit ones, latent rules must be inferred through evidence accumulation across temporal action-outcome histories.}
\label{fig:contextualGuidance}
\end{figure}

In tasks with explicit signals, rules are specified by external inputs that act as triggers, conveying immediate information about the environmental context \cite{koechlin2003architecture,driscoll2022flexible}. For example, arithmetic symbols such as $+$ or $-$ specify the required operation directly. Latent or implicit rules, where preferred or optimal actions are not explicitly signaled, must instead be inferred, e.g., via evidence accumulation or reinforcement learning (see Fig. \ref{fig:contextualGuidance}). Astrocytes may be critical to this process.  For instance, influential empirical findings \cite{mu2019glia,chen2025norepinephrine} show that astrocytes accumulate evidence of futility (i.e., unsuccessful behavioral attempts at solving the task at hand) via noradrenergic pathways. When futility crosses a threshold, animals become passive, i.e., when initial actions prove fruitless, they abandon further attempts. Separately, recent findings demonstrate that disruption of astrocytic dynamics leads to alterations in RL performance \cite{juliaPai}, providing direct evidence of astrocytic involvement in this learning pathway.  The goal of this paper is to develop and study a neural network model for astrocytic involvement in RL environments, and specifically how astrocytes may enable evidence accumulation and subsequent context-specific neuromodulation. Such mechanisms may then be leveraged toward neural-astrocyte networks within learning systems.

\subsection{Fast-slow neural-astrocyte network models}
Building on our prior modeling frameworks in \cite{gong2024astrocytes}, we formalize the computational role of astrocytes in an anatomically inspired coupled slow–fast architecture comprising a slow astrocytic network and a fast neuronal recurrent circuit. The slow network maintains astrocyte‑like variables that integrate internal and external contextual stimuli such as sensory inputs, neuromodulators, and outcome history. The modeled astrocytes then modulate the fast circuits through gain‑like, spatially structured interactions. 
We probe this architecture in a setting where context is latent and must be inferred from action-outcome observations. 

In this paper, we present an explicit state‑space characterization of the dynamical mechanisms that emerge in trained agents within this fast-slow neural-astrocyte architecture. Our results show that astrocytes mediate the accumulation of evidence regarding latent environmental parameters and accordingly modulate the faster neural dynamics for context-specific computations. The result amounts to a hybrid-automaton mechanism, manifesting as a recognizable sticky win-stay, lose-switch type policy.

\section{Model Formulation}
\subsection{Overview}
We implement a two-level hierarchical recurrent neural network with putative astrocytic modulation and separated timescales. This network operates within an RL environment in a trial-wise fashion. 
Within each trial $k$, a fast, neuron-like RNN (N-RNN) evolves its state $x^{(k)}(t)$ over $t \in \{1, \dots, T\}$ fast steps. The slow, astrocyte-like RNN (A-RNN) updates its state $h_k$ at the trial onset, based on the outcome of the previous trial $o_{k-1} = (\pi_{k-1}, r_{k-1})$. The resulting astrocytic action $\pi_k$ is held \emph{piecewise-constant} during the entire fast evolution (i.e., $t$), modulating the N-RNN’s effective dynamics. The coupled A-RNN and N-RNN dynamics are formalized as:
\begin{align}
    x^{(k)}_{t+1} &= (1-\tau_x)x^{(k)}_{t} \nonumber \\ &\qquad + \tau_x \left( \mathcal{W}_{\text{eff}}(\pi_k) \tanh(x^{(k)}_{t}) + B_{dx}u^{(k)}_{t} + \epsilon_x \right) \label{eq:fast_dyn} \\
    h_{k+1} &= (1-\tau_h)h_k \nonumber \\ &\qquad + \tau_h \left( W_{hh} \tanh(h_k) + W_{oh}o_k + b_h + \epsilon_h \right) \label{eq:slow_dyn} \\
    g_{k+1} &= \arg \max ( \nonumber \\ 
    \label{eq:action}
    &\quad \mathrm{Softmax}\left( W_2 \mathrm{ReLU}(W_1 h_{k+1} + b_1) + b_2 \right) )\\
    \pi_{k+1} &= \operatorname{OneHot}\left(g_{k+1}\right) \label{eq:policy}
\end{align}
where $g_{k+1} \in \{1, \dots, M\}$ is the discrete action index selected at trial $k+1$, and $\pi_{k+1} = \operatorname{OneHot}(g_{k+1}) \in \{0, 1\}^M$ is a standard one-hot vector with $1$ at index $g_{k+1}$ and $0$ elsewhere.

$t$ denotes the fast intra-trial time, $\tau_h \gg \tau_x$ (in terms of normalized update rates per step). The observation $o_k = (\pi_k, r_k)$ is computed at the end of trial $k$ (at $t=T$) and fed to the A-RNN to determine the modulation for the next trial $k+1$. 
The external task output is linearly decoded from the N-RNN state, via:
\begin{equation} \label{eq:lin_decoder}
y^{(k)} _{t} = W_{qx} x^{(k)} _t
\end{equation}

The N-RNN effective connectivity matrix in Eq.~\eqref{eq:fast_dyn} is defined via the Hadamard product ($\odot$) as proposed in \cite{vedovati2024synergistic}:
\begin{equation}\label{Weff}
    \mathcal{W}_{\text{eff}}(\pi_k) = J_{xx} \odot \left( \mathbf{1}\mathbf{1}^{\top} + H_{px} \operatorname{diag}(\pi_k) H_{px}^\top \right)
\end{equation}
where $\mathbf{1}\mathbf{1}^{\top} \in \mathbb{R}^{N_x \times N_x}$ is a matrix of ones with dimensions matching the recurrent matrix $J_{xx} \in \mathbb{R}^{N_x \times N_x}$. 

Through this multiplicative formulation, recurrent synaptic weights are modulated according to the active contextual state $\pi_k$. From a neurobiological perspective, multiplicative gain control reflects synaptic scaling mechanisms \cite{lefton2024norepinephrine, henneberger2010long, nagai2021behaviorally}. Because the projection matrix $H_{px} \in \mathbb{R}^{N_x \times M}$ is tall ($M \ll N_x$), the modulatory term $H_{px} \operatorname{diag}(\pi_k) H_{px}^\top$ induces a low-rank transformation. This is intended to capture the notion of astrocytic tiling, wherein astrocytic and neuromodulatory gating targets localized neuronal ensembles rather than uniform global networks \cite{nadim2014neuromodulation, papouin2017astrocytic}.

\subsection{Task and environment set-up} \label{subsec:env}

\begin{figure}[b]
    \centering
    \includegraphics[width=0.95\linewidth]{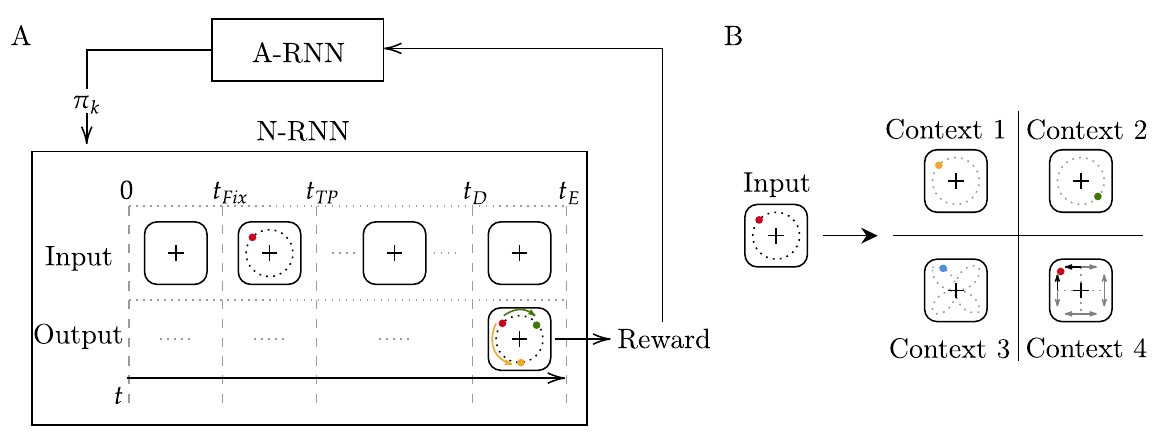}
    \caption{Hierarchical task schematic. A. Timeline of a single delayed match-to-sample trial executed by the N-RNN. After a brief fixation span, the network receives a positional stimulus during input presentation from $t_{\text{Fix}}$ to $t_{\text{TP}}$ and holds the memory across a variable delay period until $t_{\text{D}}$, when a following a trigger, is required to reach the target within a response window $t_{\text{E}}$. B. Representation of the $M=N=4$ contexts mapping.}
    \label{fig:task_schematic}
\end{figure}
    
The coupled system solves a hierarchical task comprising two interconnected levels: a deterministic low-level Romo-like task and a high-level stochastic bandit environment \cite{romo1999neuronal}, schematized in Fig.~\ref{fig:task_schematic}. 
At the lower level, the N-RNN interacts with a delayed match-to-sample task (on the time-scale of $t$), in which, in each trial $k$, the agent processes an external stimulus representing a point on a circle. Its objective is to retain the given information over a random temporal delay and, when instructed, reach a corresponding target. This will derive a reward (see Appendix~\ref{app:task}). The correct stimulus-target mapping is dictated by the active latent context $c_k \in \{1, \dots, M\}$ (here, $M=4$). Correctly executing the low-level task thus requires inference of the context. The N-RNN is modulated by the astrocytic network, which emits a contextual signal $\pi_k \in \{e_1, \dots, e_M\}, e_i \in \mathbb{R}^M$ that gates the synaptic weights in Eq.~\ref{Weff}. The astrocytic network has access to the rewards derived from the correct execution of the low-level task. 

The task environment maintains an active context for a variable block duration of $L$ consecutive trials, where $L \sim U(3, 25)$. At the end of each block, the environment transitions to a context chosen uniformly at random from all $M$ available contexts.

Reward delivery is also stochastic. When the correct context is selected and the N-RNN successfully executes the low-level task, a reward is delivered with probability $p^\star$. Any other context that is (incorrectly) emitted yields a reward with probability $(1-p^\star)/(M-1)$. This formulation ensures that individual reward outcomes carry high variance and limited instantaneous information. A zero-reward outcome ($r_k=0$) is inherently ambiguous because it might signal either a context shift or a probabilistic reward omission under a correct context.

In sum, the astrocyte network is performing an M-arm bandit problem, selecting the context to emit and receiving an ensuing stochastic reward from which to determine future emissions. In the bandit/RL parlance, these emissions are actions of the astrocytic networks and we will use these terms interchangeably. We consider four entropy regimes. 

\begin{enumerate}
    \item Low stochasticity ($H = 0.627$): reward feedback is strongly predictive of subsequent outcomes, with the correct action yielding a reward with probability $p^\star = 0.90$.
    
    \item Moderate stochasticity ($H = 1.208$): reward feedback becomes less predictive of subsequent outcomes, as the reward probability for the correct action decreases to $p^\star = 0.75$.
    
    \item High stochasticity ($H = 1.604$): the predictive value of reward feedback declines further, with the correct action rewarded with probability $p^\star = 0.60$.
    
    \item Maximum stochasticity ($H = 1.838$): reward feedback is only weakly predictive of subsequent outcomes, as the reward probability for the correct action falls to $p^\star = 0.48$.
\end{enumerate}

\subsection{Model training}

We train the coupled architecture by decoupling the optimization of the two subnetworks to mirror their distinct operational roles and temporal scales. The low-level N-RNN is trained via supervised learning to minimize a mean squared error (MSE) loss with respect to the target reach location. 

To train the A-RNN, we note that a per-trial supervised loss does not reliably reveal the underlying latent context in a stochastic environment \cite{behrens2007learning, angela2005uncertainty}. 
Consequently, we train the A-RNN using reinforcement learning and opt for Advantage Actor-Critic with Generalized Advantage Estimation (A2C+GAE) \cite{chen2021adaptive}, evaluated on a sparse, one-bit trial reward. This paradigm allows the model to shape its processing of the action-outcome observation stream across multiple trials. The resulting trained dynamics implement a sticky win-stay/lose-switch scheme, revealing a dynamical mechanism by which astrocytes can support contextual switching in neural circuits.

To accelerate early convergence and stabilize value function estimation, we employ a curriculum schedule: during the first 500 trials, the reward magnitude for correct contextual matches is temporarily augmented ($r_k = 10$), followed by a brief 250-trial annealing phase where the reward scale is stepped down incrementally to unit magnitude ($r_k \to 1$). During this warm-up, the coupled system relies solely on the A-RNN's continuous state dynamics to infer the latent context and guide task execution (see also Appendix~\ref{app:sched}).

\subsection{Relationship to Meta-RL and Hypernetworks}

Our work adopts training paradigms similar to those used in meta-reinforcement learning \cite{wang2016learning, duan2016rl, wang2018prefrontal}. However, our goals are different in the following sense. Meta-RL seeks an agent capable of adapting its policy based on interactions with a dynamic environment. Our trained models do not adapt \textit{per se}, but embed a fixed policy that responds to environmental inputs.  Our focus is on how the architectural constraints of the neural-astrocyte hierarchy enable the emergence of such a policy.

Furthermore, our approach departs from conventional context-based or multi-task RL methods \cite{humplik2019meta, rakelly2019efficient, zintgraf2019varibad}. Many multi-task and multi-context frameworks assume that the context variable is either explicitly provided or sampled from an oracle. In contrast, the slow A-RNN autonomously infers latent contextual shifts by processing its own historical action-reward trajectories $o_k = (\pi_k, r_k)$ across consecutive trials. Additionally, unlike systems that periodically reset hidden states to zero at task boundaries, the A-RNN evolves continuously across trial blocks, allowing long-term history to shape its state-space trajectory.

Finally, our model differs in how the contextual signal is delivered to the task network. Standard context-based methods concatenate the inferred context directly onto the input signal \cite{driscoll2022flexible, yang2019task}. Instead, we implement a form of dynamic parameter selection akin to \emph{hypernetworks} \cite{ha2016hypernetworks}, where the slow A-RNN reconfigures the fast N-RNN's connectivity matrix $\mathcal{W}_{\text{eff}}(\pi_k)$ on a trial-by-trial basis.

\section{Results}
\subsection{Networks learn an automaton-like sticky policy}

\begin{figure}
    \centering
    \includegraphics[width=\linewidth]{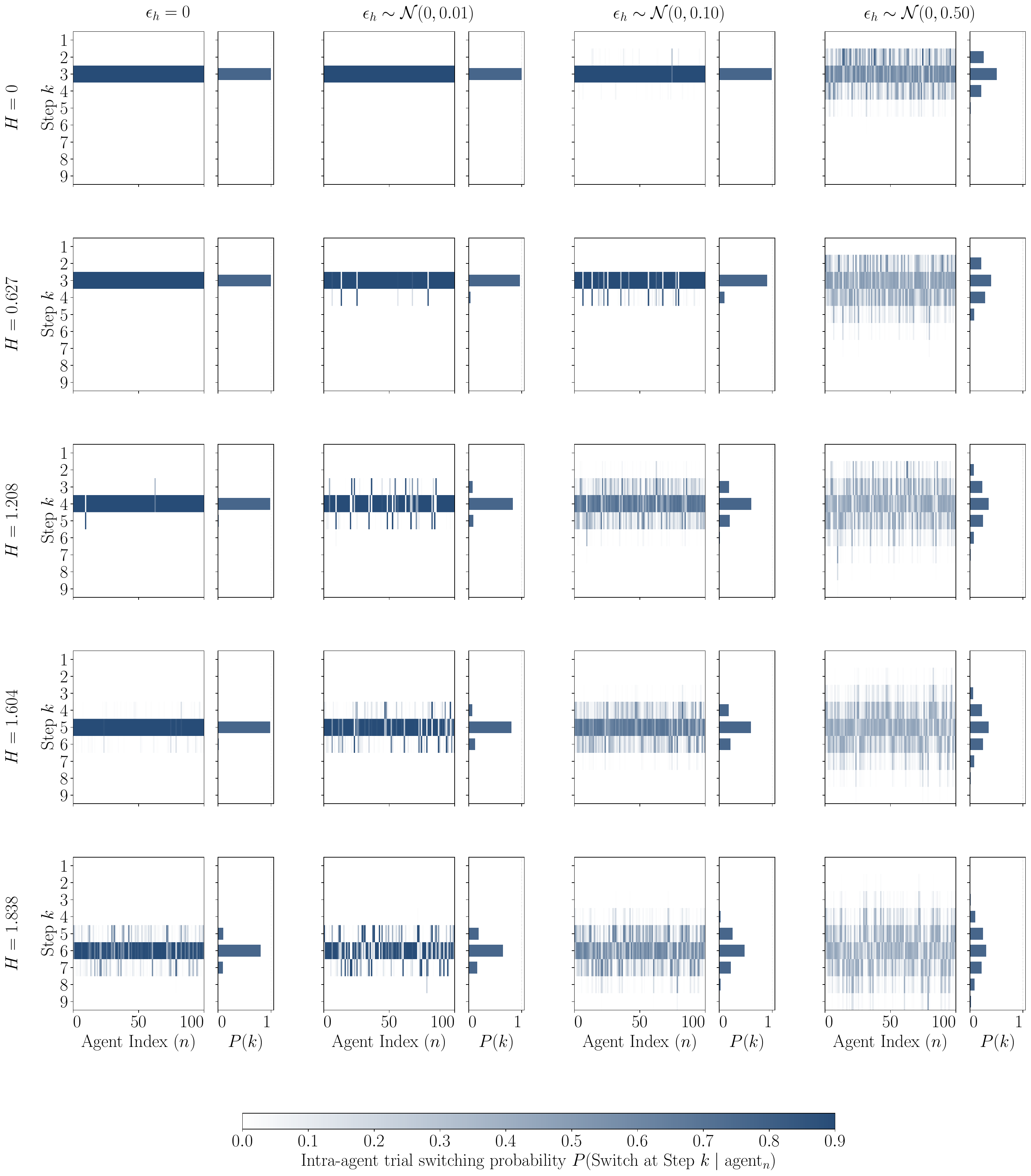}
    \caption{Learned context-switching latencies across environmental entropy levels ($H$) and internal noise magnitudes ($\epsilon_h$). Each plot illustrates the empirical intra-agent trial switching probability $P(\text{Switch at Step } k \mid \text{agent}_n)$ for 100 trained models ($n=1\dots100$), as a function of successive unrewarded trials, alongside the corresponding marginal distribution.
    At low entropy, switching occurs almost deterministically. As environmental entropy increases, the average stickiness increases as does the temporal variance. Increasing internal noise ($\epsilon_h$) also broadens the marginal distributions $P(k)$.
    }
    \label{fig:k_step_transition}
\end{figure}

We proceed to study the dynamics of the astrocyte-neural network within the specified task setting. At the heart of our analysis is the modulation of neural activity by means of the astrocytes, i.e., the evolution of $\pi_k$ through the course of trials and ensuing effect on $\mathcal{W}_{\text{eff}}$ via \eqref{Weff}. 
We trained 100 independent models under fixed environmental entropy and external noise $\epsilon \sim \mathcal{N}(0, 0.01)$, initialized with distinct random weights (see Appendix~\ref{app:params}). Following convergence, we evaluate the models across a spectrum of internal noise levels ($\epsilon_h$). 
Intuitively, in a low-entropy environment, a behavioral strategy aligns with a strategy akin to `win-stay, lose-switch,' in which the first occurrence of a reward-prediction error (RPE) triggers an immediate action switch. On the other hand, in a high-entropy scenario, an agent benefits from sticking with a choice across successive RPEs before committing to a switch.

\begin{figure}
    \centering
    \includegraphics[width=\linewidth]{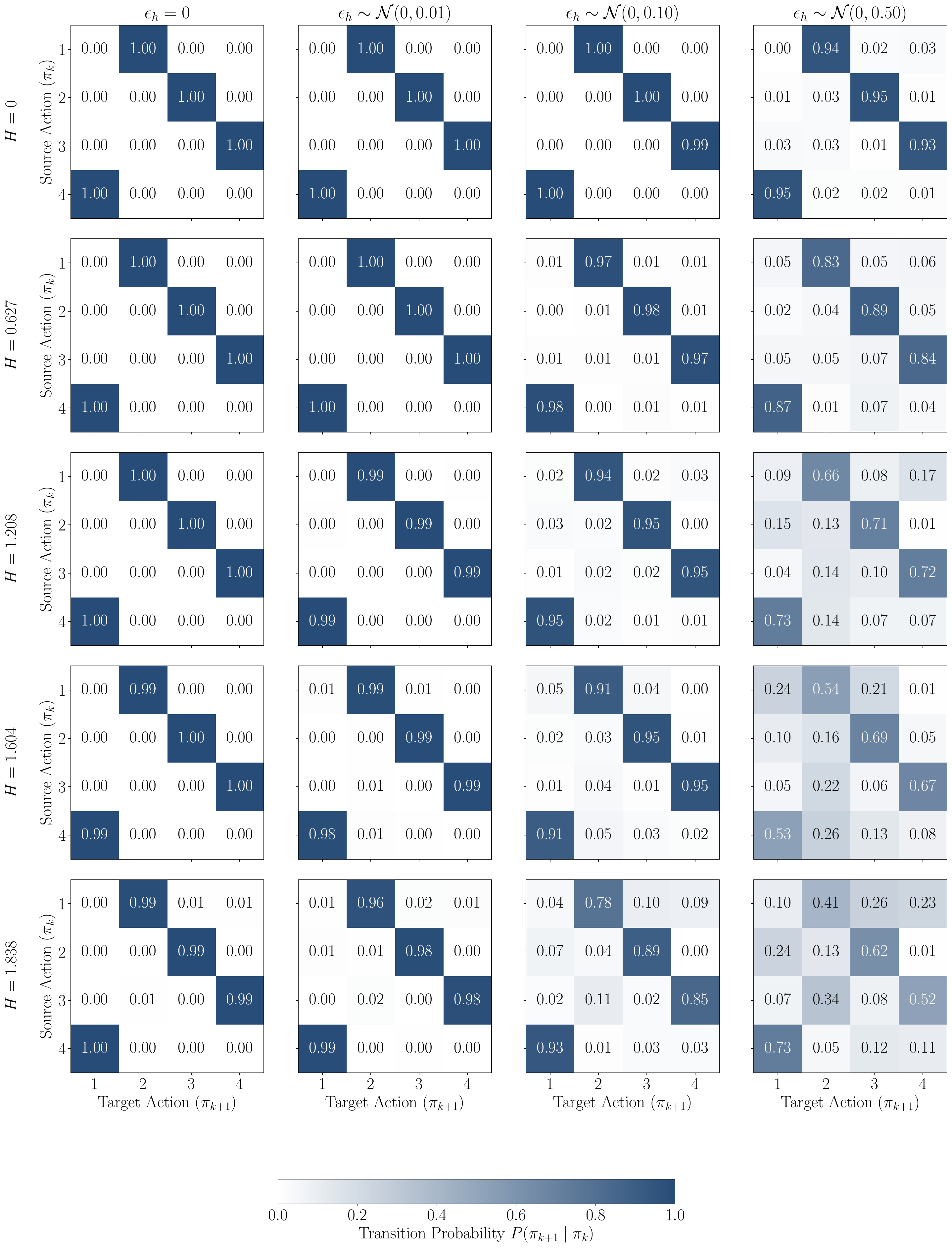}
    \caption{Empirical action transition probability matrices $P(\pi_{k+1} \mid \pi_{k})$ associated with action switches across varying environmental entropy ($H$) and internal astrocytic noise levels ($\epsilon_h$). Each heatmap displays the probability of transitioning from a source action $\pi_{k}$ (y-axis) to a target action $\pi_{k+1}$ (x-axis) following a context switch across 100 independently trained models. Networks maintain a robust, directional policy across all parameter regimes. Under maximum environmental entropy ($H = 1.838$) and heavy internal perturbation ($\epsilon_h = 0.50$), the dominant probabilities start to fade, while still showing a tendency to remain focused along the target off-diagonals.}
    \label{fig:probability_resolution}
\end{figure}

To corroborate our hypothesis, we examine two interconnected aspects of the agent's learned behavior across varying levels of environmental entropy ($H$) and noise ($\epsilon_h$).
First, we evaluate the time-to-switch distribution (Figure~\ref{fig:k_step_transition}). This measure quantifies the number of cumulative RPEs required before an agent changes its action choice. Our simulations confirm the hypothesized trend: contextual action switches occur after an entropy-dependent number of unrewarded trials (i.e., stickiness).

The second aspect of interest is the heterogeneity of policy adherence across the agent population, which we probe by computing transition probabilities associated with action switches (Figure~\ref{fig:probability_resolution}). To obtain this figure, we standardized the action space by defining $a_1, \dots, a_4$ as the first four actions taken for any given network. 
The observed behavior is suggestive of a hybrid automaton: actions are switched according to a set policy (here, cyclic), while an inner dynamics governs \textit{when} such transitions are triggered (i.e., the step-wise stickiness shown in Figure~\ref{fig:k_step_transition}). Of note, the behavior we observe here is not purely deterministic. While the policy exhibits a dominant cycling structure, small behavioral deviations occur, particularly at higher entropy levels. Some of this dispersion is undoubtedly driven by the astrocytic noise level $\epsilon_h$. To probe this effect, we repeated the analysis across variable noise regimes, observing increases in both stickiness and cycling dispersion as noise scales increased. Interestingly, however, some structural variation persists even in the completely noise-free model ($\epsilon_h = 0$), particularly under elevated entropy conditions.

\subsection{Rule switches are mediated through action-reward bifurcations}

\begin{figure}
    \centering
    \includegraphics[width=0.95\columnwidth]{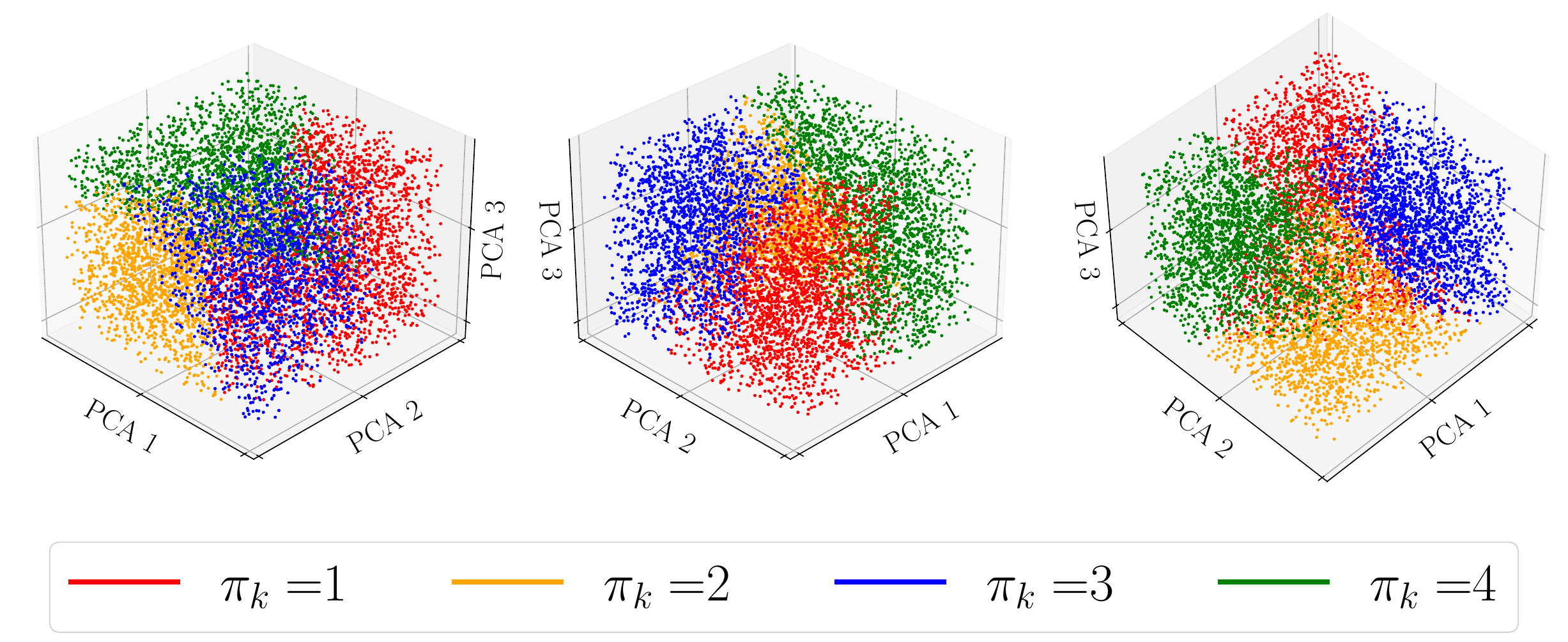}
    \caption{Low-dimensional state-space projections showing the geometric partitioning of the astrocytic network into distinct action-level sets, viewed from three different rotational perspectives.}
    \label{fig:level_sets}
\end{figure}

\begin{figure*}
        \centering
        \includegraphics[width=0.85\linewidth]{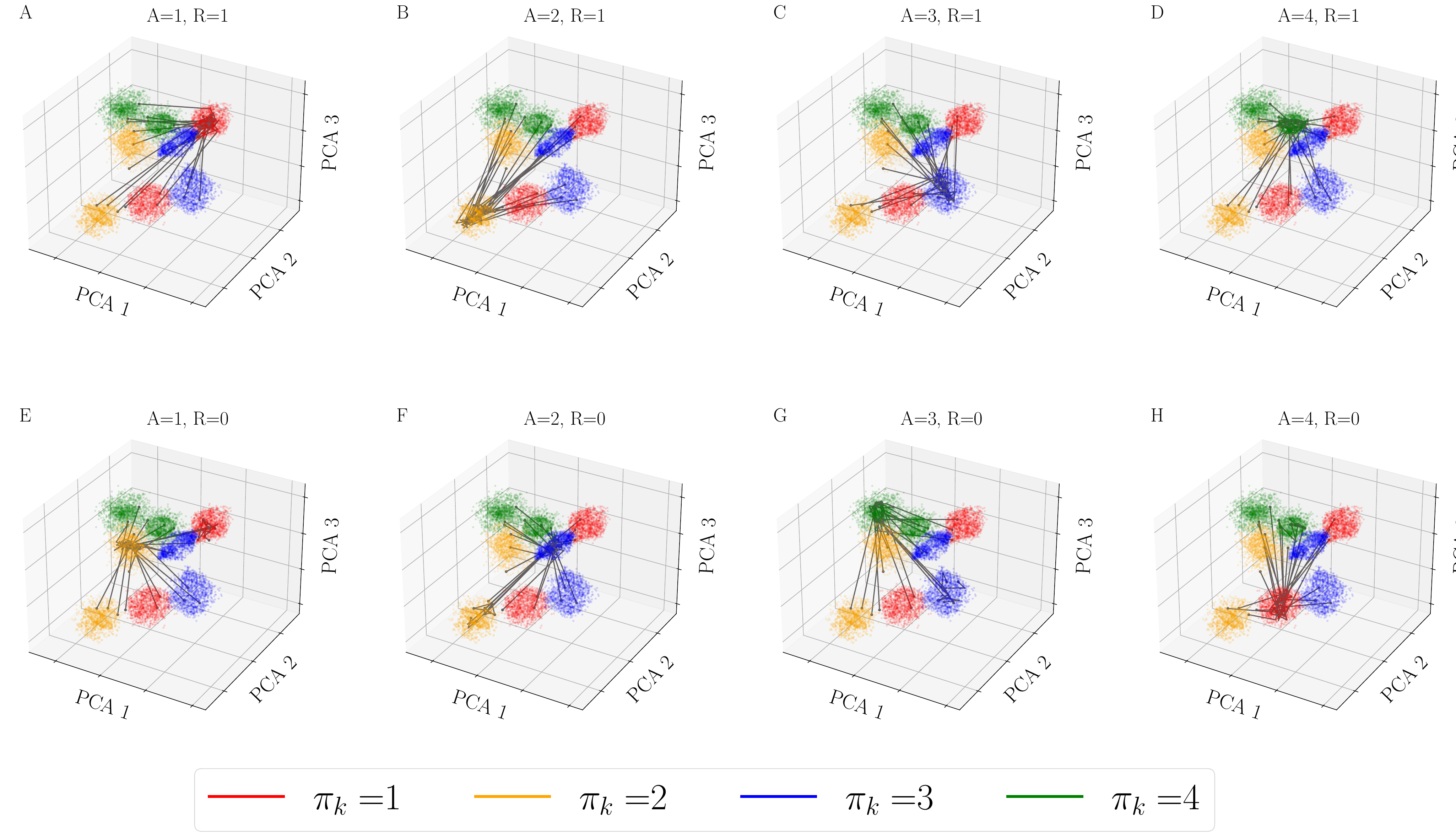}
        \caption{Autonomous state-space trajectories under fixed action ($A$) and reward ($R$) inputs. (top): $R=1$ showing consolidation of the active context, (bottom) $R=0$, showing traversal of context.}
        \label{fig:trajectories_}
\end{figure*}

To better understand this behavior, we shifted our attention to a deeper analysis of the model dynamics and internal mechanisms. 
To begin the analysis, we first recognize that the distinct decoded contexts $g_k$ are, in fact, level sets of \eqref{eq:action}, such that action $c \in {1,\dots, M}$ is emitted for all network states satisfying:
\begin{equation}
\arg \max (\mathrm{Softmax}\left( W_2 \mathrm{ReLU}(W_1 h_{k} + b_1) + b_2 \right)) = c.
\end{equation}
We note that these level sets are independent of the emitted action or derived reward. Figure~\ref{fig:level_sets} depicts four level sets for an illustrative low-entropy example. Our analysis then centers on how the dynamics mediate the traversal of these level sets, corresponding to context switching.

To analyze the dynamics relative to these level sets, we first note that the vector field (i.e., \eqref{eq:slow_dyn}) \textit{is} affected by the action and reward, such that the action and reward act as bifurcation parameters that alter the dynamics.  Figure~\ref{fig:trajectories_} shows the flow of several trajectories for the above example, sampled across all four level sets. What we see is that each action-reward pair is associated with a distinct limit set (in this case, a limit cycle attractor), located within one of the action level sets.  Thus, when exogenously “forced" to enact a specific action and receive a specific reward, the network evolves to a limit set associated with a corresponding (new) action. 
Interestingly, each action level set contains two such limit sets; one is associated with successive reward, versus the other associated with successive unrewarded trials.

\subsubsection{Dynamics and contextual inference}

To study how these different attractors are used in the service of contextual inference, we examine the trajectories in the vicinity of each action-specific level set. As illustrated in Figure~\ref{fig:trajectories_}, there are essentially two scenarios that can occur. First, the network can be configured to emit an action that is subsequently rewarded (Fig.~\ref{fig:trajectories_}, top row).  In this case, those actions are, in essence, reinforced, and trajectories converge to the limit cycle attractor within that portion of the action level set.  We refer to this as a rewarded attractor.

On the other hand, those same actions could be subsequently unrewarded.  In this case, trajectories eventually approach a limit set in a different action level set (Fig.~\ref{fig:trajectories_}, bottom row), causing the emitted action to change in subsequent trials. In the illustration, this traversal takes three steps (i.e., the network sticks with the original action three times), though this varies according to the entropy of the bandit environment, as we elaborate next.

\subsubsection{Vector field geometry and stickiness}

\begin{figure}[!t]
\centering
\includegraphics[width=\columnwidth]{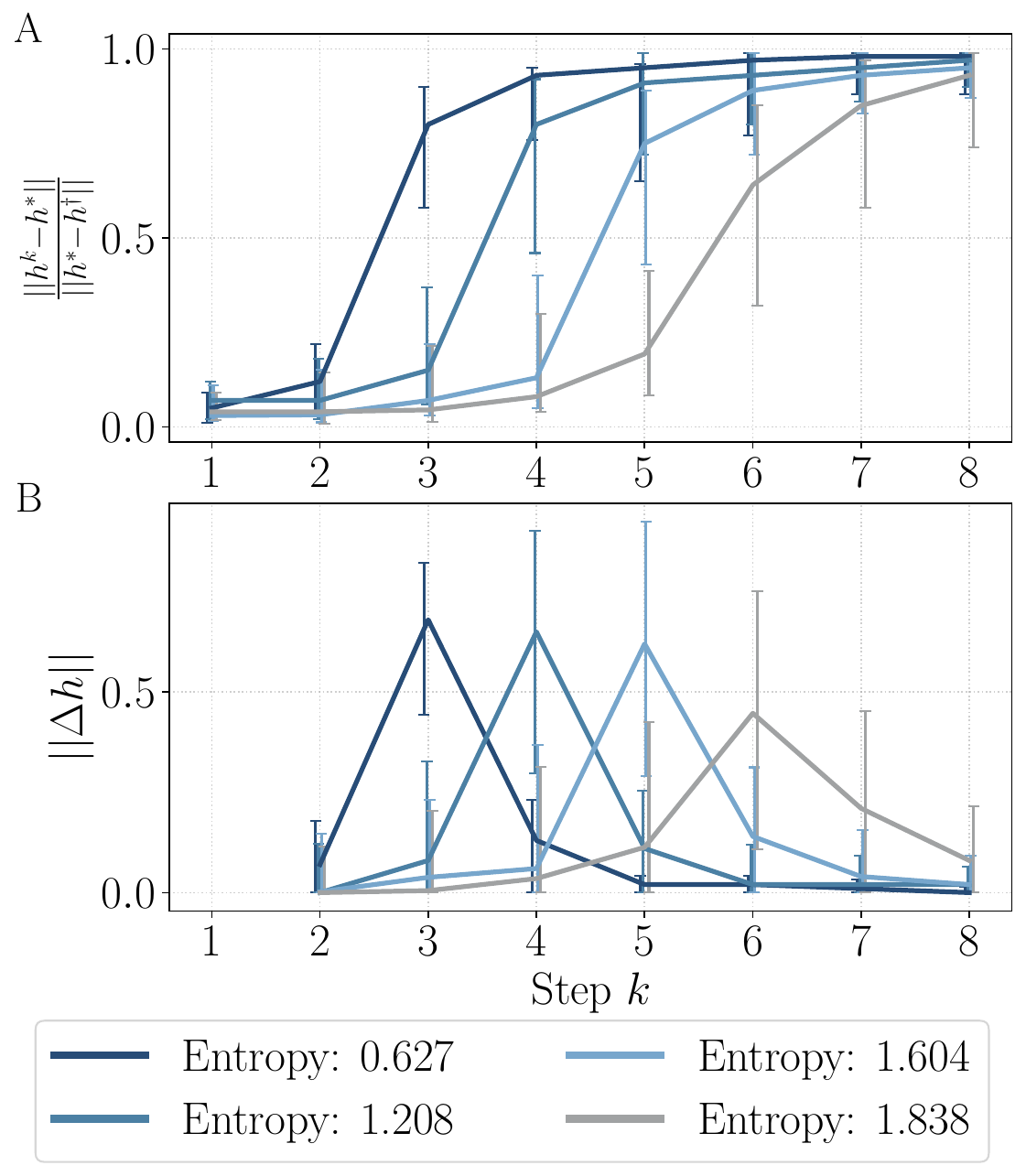}
\caption{Normalized distance and instantaneous velocity from rewarded to unrewarded conditions. Note that $h^{*}$ and $h^\dagger$ represent the empirical centroids of the initial ($R=1$) and target unrewarded attractor sets, respectively.}
\label{fig:stickiness_mech}
\end{figure}

The stickiness property of the different agents shows the magnitude of the vector field under unrewarded conditions. To show this, we collected data from 100 trained agents and quantified unrewarded trajectories starting with the first encounter of an RPE. 
Figure~\ref{fig:stickiness_mech} depicts two quantities of interest: (i) the normalized distance from a prior rewarded attractor to an ensuing (upon successive RPEs) unrewarded attractor, and (ii) the speed achieved during the traversal. These quantities are shown for each different level of environmental entropy.

\begin{figure}
    \centering
    \includegraphics[width=0.95\columnwidth]{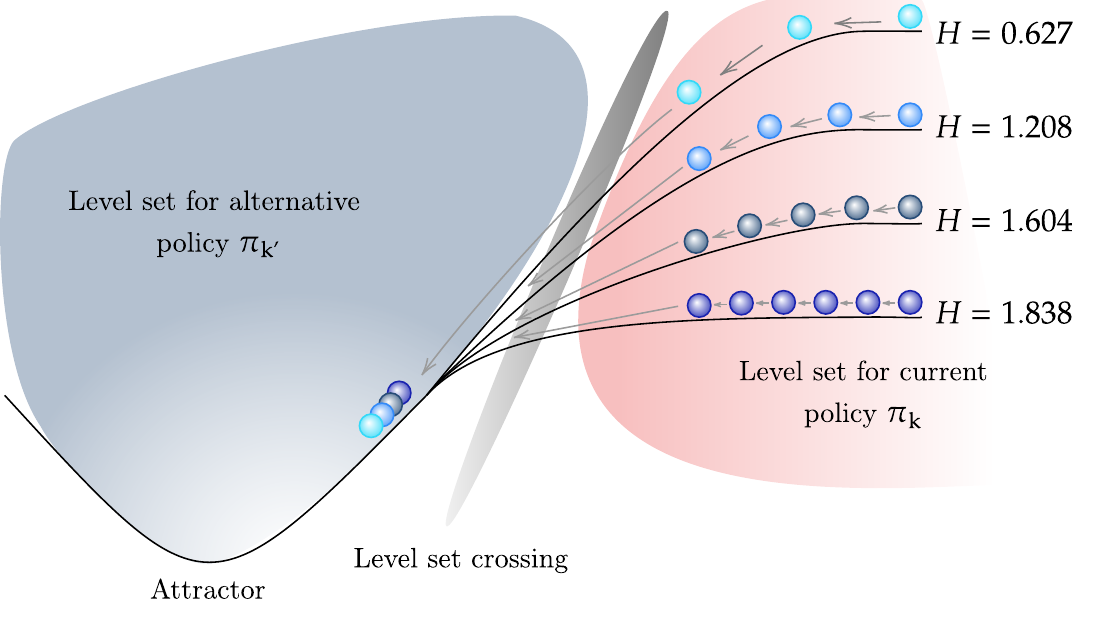}
    \caption{Conceptual schematic of autonomous trajectories traversing action level sets under the unrewarded vector field $(\pi_k, r_k=0)$. Upon reward omission, a global bifurcation annihilates the active attractor and reconfigures the vector field. The trajectory evolves autonomously through a low-velocity region within the current context level set ($\arg \max \text{Softmax}(h) = c_k$, red region). Higher environmental entropy flattens the local landscape, increasing the number of discrete steps required to cross the decision boundary (grey threshold). Once the trajectory crosses this boundary into the alternative action level set ($\arg \max \text{Softmax}(h) = c_{k'}$, blue region), the network undergoes a second rapid vector field change toward the new attractor well, cleanly emitting the alternative action mode.}
    \label{fig:vector_field_scheme}
\end{figure}

To interpret these data, it is useful to assume that the model has maintained a consistent action-outcome history $(\pi_k,r_k=1)$ for some number of steps prior to the RPE.  In such a condition, the trajectory will have converged to an asymptotically stable limit cycle attractor (i.e., returning to Figure \ref{fig:trajectories_} (top)). Upon the RPE, the system bifurcates, restructuring the vector field such that the previous limit cycle is annihilated and a new target limit set is established in a different region of the state space (again, see Figure \ref{fig:trajectories_} (bottom)).  Figure~\ref{fig:stickiness_mech} shows that the stickiness of $\pi_k$ corresponds to the steepness of the vector field proximal to the previous $(\pi_{k-1},r_k=1)$ attractor.  
The overall picture, then, is schematized in Figure~\ref{fig:vector_field_scheme}, which portrays the steps the agents take to cross level sets. Under high entropy, the unrewarded vector field is flatter, requiring more time-steps to advance toward the new level set attractor.

This geometric framework also explains the variability observed in the models at different noise levels discussed previously. Even in the absence of external noise, stochasticity in the feedback rewards remains a critical driver. In higher-entropy scenarios, when level set boundaries are flatter, such randomness can place the network's trajectories in areas of ambiguity where two or more level sets interface, leading to occasional cycling and timing variation (see again Figure ~\ref{fig:probability_resolution}). 

\begin{figure}
    \centering
    \includegraphics[width=0.85\columnwidth]{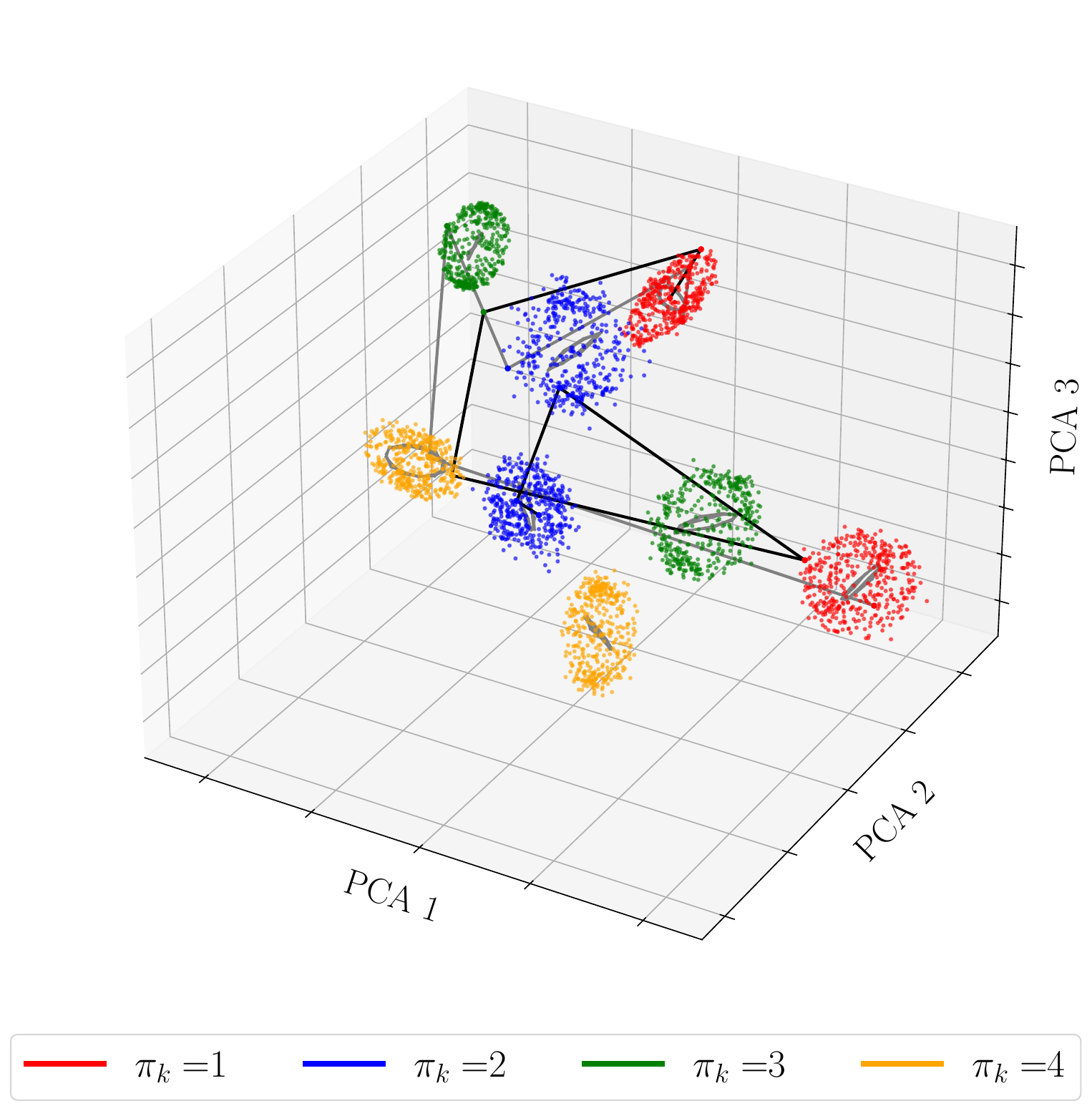}
    \caption{Visualization of a missed cycle. The light gray line represents the standard action selections, while the black line represents an anomalous one.}
    \label{fig:missed_cycle}
\end{figure}

To isolate the geometric mechanism driving these policy variations, we analyzed a representative single-trial state-space trajectory (Figure~\ref{fig:missed_cycle}), in which the nominal trajectories under standard evolution conditions are shown in light gray. We highlight in black an anomalous trajectory triggered by an unrewarded trial ($R=0$). The network exhibits action stickiness, remaining in the red level set for an extra trial. Upon the subsequent step, the continuous state vector would be nominally expected to traverse into the ($\pi_k = 2$, blue) level set. However, the actual trajectory ends up just across the boundary into the domain where $\pi_k = 3$ (green) is decoded. 

\subsection{Astrocytic tiling and lattice topology are sufficient for network-wide modulation}

\begin{figure}[!t]
    \centering
    \includegraphics[width=0.95\columnwidth]{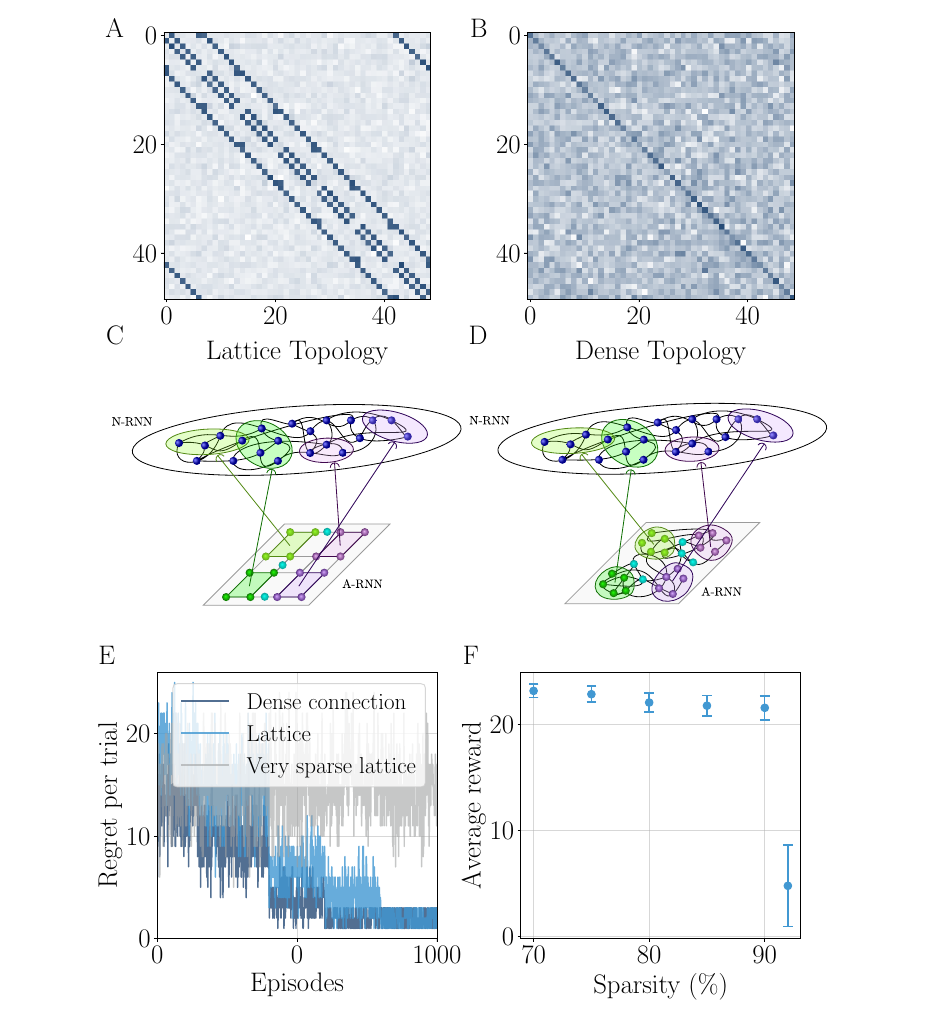}
    \caption{Lattice-constrained astrocytic recurrence matches dense performance.}
    \label{fig:lattice_perf}
\end{figure}

Our model presumes a certain level of connectivity between astrocytes.
Biologically, it is known that astrocytes form a sparse, gap-junction mediated network \cite{cooper2026astrocytes}, with each astrocyte overlaying a spatially contiguous domain of neurons (known as `tiling', e.g., \cite{hayashi2022neurons}).
In this work, we chose to mimic this topology and assumed a 1:3 astrocyte-to-neuron ratio. We organized astrocytes into non-overlapping spatial domains, forming a tiled structure similar to a lattice topology. We asked whether this assumption limits the dynamical repertoire or behavior exhibited by the trained models, since canonical RNNs typically rely on dense, all-to-all recurrent connectivity.

In the lattice topology, each astrocyte interacts only with its $k$ nearest neighbors \cite{genoud2015proximity,bushong2004maturation}, yielding a banded recurrent connectivity (Fig.\ref{fig:lattice_perf}A) in contrast to the dense baseline (Fig.\ref{fig:lattice_perf}B). Across training runs, imposing lattice topology has little effect on dynamics or performance: lattice-coupled models train at a comparable rate and ultimately reach similar asymptotic regret as the fully connected models (Fig.~\ref{fig:lattice_perf}E). This result supports the idea that effective network-wide context integration can emerge solely from local astrocytic interactions. However, this locally mediated integration finds a limit once connectivity becomes too sparse to support reliable communication across the network. 
As shown in Fig.~\ref{fig:lattice_perf}F, average reward remains high across a broad range of sparsity (proportion of nonzero entries) levels, up to $ 90\%$.

\subsection{Hierarchical timescale separation advantages trainability}

\begin{figure}[!t]
\centering
\includegraphics[width=0.95\columnwidth]{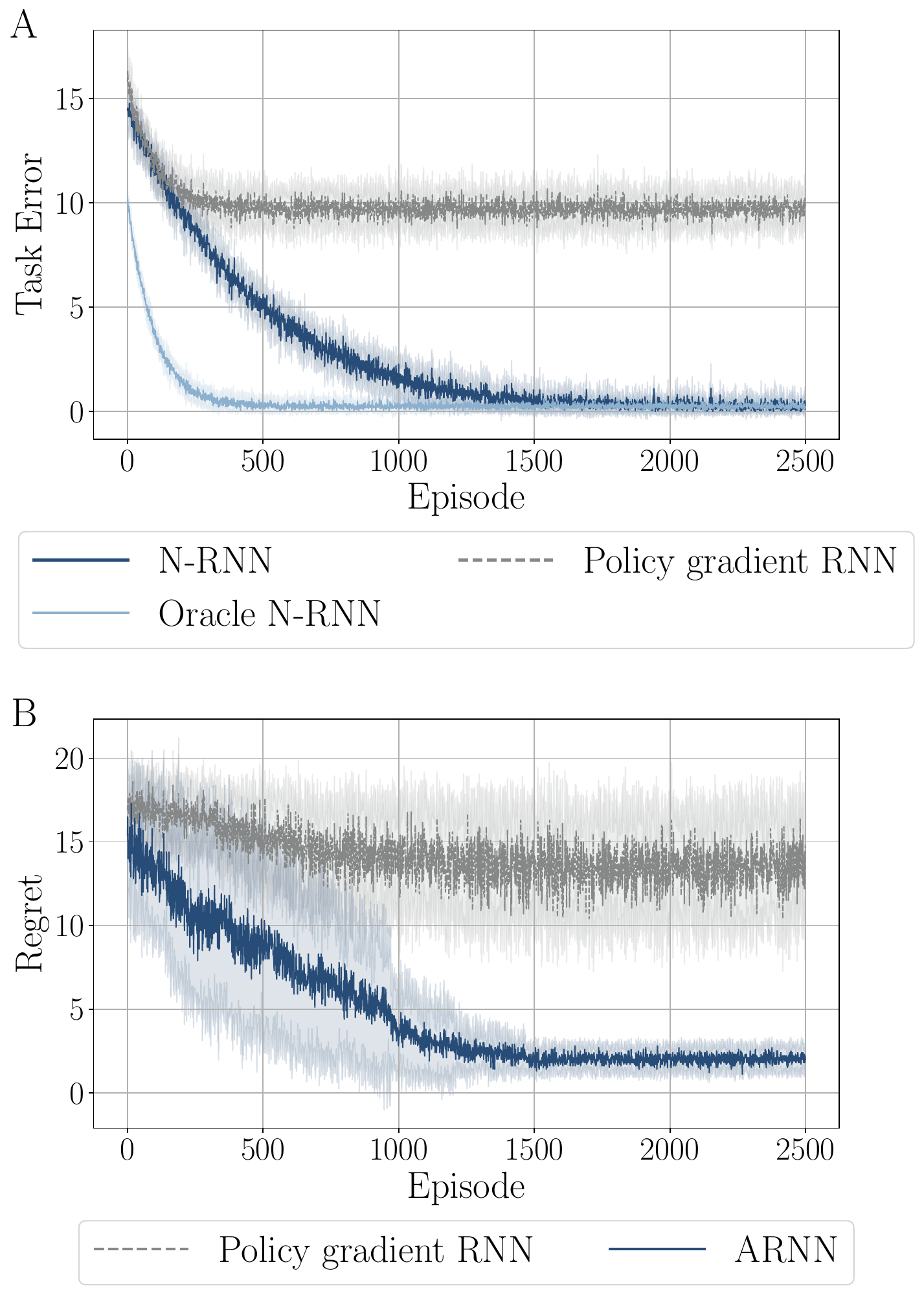}
\caption{Convergence results: comparing the hierarchical models with other architectures.}
\label{fig:summary_learning_dyn}
\end{figure}

A key premise of this work is that the time-scale separation of astrocytes and neurons provides a natural and computationally efficient way to hierarchically separate contextual evidence accumulation from low-level task execution. We posited that this architecture would aid the trainability of neural-astrocyte networks versus conventional RNNs with the same number of units, but without hierarchical structure or time-scales.

To study this, we established an oracle baseline for the task by training a supervised RNN that is provided with explicit ground-truth contextual inputs. This oracle defines the performance ceiling achievable when top-level context selection is fully known and deterministic.

Second, we evaluated a conventional policy-gradient RNN equipped with dual readout heads, one dedicated to the task output and the other to continuous context identification. This network received task inputs, the previous action identifier, and the corresponding scalar reward signal. Rewards were distributed stepwise based on task accuracy, along with a trial-end bonus for correct context identification. In essence, this is a monolithic RNN of the same size (number of parameters) as our neural-astrocyte model.

As demonstrated in Figure~\ref{fig:summary_learning_dyn}, the monolithic architecture struggles to learn the task, even though, in principle the neural-astrocyte model is a special case of the monolithic architecture.  
We attribute this result to the premise that a single recurrent population must simultaneously preserve persistent, low-frequency reward histories for context inference and generate high-frequency transient dynamics for within-trial requirements. The high-variance policy-gradient updates driving fast execution may thus overwrite the slow credit assignment required for latent-state estimation, preventing convergence. The neural-astrocyte model resolves this issue by explicitly decoupling timescales, assigning continuous evidence integration to the slow top-level network (A-RNN), which provides a stable modulatory signal to the fast network (N-RNN).

\section{Conclusions}
This work presents a computational framework that formalizes astrocytes as contextual switchboards within neural networks, suggesting how timescale separation between slow astrocytic information integration and rapid neuronal execution constitutes an integrative and flexible computational substrate.

\subsection{Bifurcations and a hybrid-automaton mechanism}

Our modeling demonstrates a mechanism by which a dynamical neural network can achieve an analog of discrete state-transition rules in classical automata theory. In our model, these transitions are achieved via a series of bifurcations, driven by RPEs.

When the agent makes a valid contextual selection $c$ under rewarded conditions, the system operates under a vector field parameterized by the tuple $(\pi_k = \operatorname{OneHot}(c), r_k = 1)$, where the network state settles into a stable, attracting limit cycle. Upon reward omission, the parameter update triggers a global bifurcation, reconfiguring the phase space into an unrewarded vector field parameterized by $(\pi_k = \operatorname{OneHot}(c), r_k = 0)$ and yielding the low-velocity dynamical signatures described previously.

Crucially, completing a full context switch relies on a sequential cascade of two distinct bifurcations. 
First, the reward omission ($r_k=0$) triggers a bifurcation that annihilates the active attractor. The state vector $h_k$ then flows through a flat potential region within the current contextual domain $\arg \max \mathrm{Softmax} (h) = c$. As $h_k$ continues along this trajectory, it crosses the boundary into a new decision region $\arg \max \mathrm{Softmax} (h) = c^{'}$, causing the readout layer to decode a new action. This update induces a second bifurcation, reconfiguring the vector field around the new context, upon which the process resets. 


In sum, the astrocytic dynamics implement a mechanism that integrates successive RPEs, or evidence of contextual incongruence. When evidence reaches a sufficient level, a switch (i.e., level set traversal) occurs, altering the low-level neural dynamics. We can interpret such a mechanism as a generalization of classical integrate-to-bound models, where the bifurcations change the location of (multidimensional) bounds in state space.  The combined network thus segregates latent-state inference from immediate action execution, wherein astrocytes provide the slow temporal substrate required to preserve contextual representations across extended trials.

\subsection{Topological robustness of the astrocytic network}

Our network suggests computational efficiencies associated with neural-astrocyte biology. Our model embeds local diffusive communication and spatial tiling, properties of biological astrocytes. In biological tissue, astrocytes are not coupled through direct, all-to-all contact but through local gap-junction–mediated coupling and diffusive signaling pathways such as calcium waves and ATP \cite{ giaume2010astroglial, mu2019glia}. Our results suggest that this decentralized astrocytic architecture is a computationally efficient solution for achieving broad modulation of neuronal dynamics. Interestingly, astrocytic modulation continues to be efficacious up to $\sim 90\%$ connection sparsity, suggesting that astrocytic networks operate in a regime of robustness without fragility. Astrocytes tolerate substantial heterogeneity and loss of local connections while maintaining global function, similar to premises regarding astrocytic connectivity in \cite{kozachkov2025neuron}. Thus, astrocytic tiling of neural networks may reflect an efficient solution to the network-theoretic challenge of achieving global integration through local communication.

\subsection{Limitations and Future Directions}
While the neuro-astrocytic framework demonstrates how astrocytic evidence accumulation can support contextual switching via a hybrid automaton, its architectural constraints also point to clear directions for future research. In our formulation, the lengths of blocks during which the context is invariant are stochastically sampled ($\sim U(3, 25)$ trials). This exposes the network to a distribution of volatility regimes. It is known that stickiness in animal choice behavior may be directly associated with volatility \cite{monosov2020outcome}, and future work may tackle this question by parameterizing the switch rates.
Currently, our hierarchical architecture is optimized over a parameterized task family and evaluated on instances drawn from the same parametrization. The slow A-RNN successfully infers latent rule shifts within this task space. However, generalizing a single pre-trained agent across structurally distinct task regimes may require alternative meta-learning procedures \cite{wang2016learning,duan2016rl,wang2018prefrontal}.

\section*{Acknowledgment}
This work was supported by grants W911NF2110312 from the US Department of Defense and 2424096 from the US National Science Foundation.

\bibliographystyle{ieeetr}
\bibliography{references}

\appendix

\section{Appendix A: Task, Implementation Details, and Training Paradigm}

\subsection{Task setup} \label{app:task}

The low-level N-RNN interacts with a spatial working memory task, similar to the one implemented in \cite{mante2013context}, and is diagrammed in Figure~\ref{fig:task_schematic}-A.
The task relies on a stimulus consisting of a 3-dimensional vector that encodes the Cartesian coordinates of a point on the unit circle. The first two components represent the point's location, while the third component encodes a response trigger. A task trial is comprised of multiple timesteps and can be divided into four distinct phases:

\begin{enumerate}

\item Fixation. The network's output is maintained at a neutral central point, while a constant null signal $\left[0,0,0\right]$ is provided to the network.  

\item Target Presentation. During this phase, the network receives information about the position of a point on a circular path, which encodes the spatial information the network must retain in memory. During this phase, the network is required not to produce any output.

\item Variable Delay Period. During this part of the task, the network’s output must once again remain neutral. At this time, the network needs to retain information about the target position for a variable time duration. This phase evaluates the network's memory capacity, as it must retain access to the stimulus information in the absence of additional inputs.

\item Response Phase. In this final phase, the network is required to produce the correct output when the input's third component is 1 for a fixed duration.

\end{enumerate}

Within this task structure, the contextual selection made by the astrocytic network via \(\pi_k\) establishes a specific relationship between the network's input and the desired output. In context one, the network reproduces the identical input presented during the target presentation phase. Context two involves a transformation in which the input is shifted 180 degrees. Context three applies the mapping defined by the function \(f(x, y) = (x \log(1 + y), y \log(1 + x))\). Lastly, context four projects points within the quadrant according to the following piecewise function: \[
f(x, y) =
\begin{cases}
x = 1 , \, \text{if } x \ge 0, \, \, x = -1, \, \text{if } x < 1, \\
y = 1 , \, \text{if } y \ge 0, \, \, y = -1, \, \text{if } y < 1, \\
\end{cases}
\]

\subsection{Parameter Initialization Specifications} \label{app:params}

\begin{table}[h!]
\centering
\caption{\textbf{Parameter initialization specifications.}}
\label{tab:initializations}
\begin{tabular}{lll}
\hline
\textbf{Parameter / Matrix} & \textbf{Symbol} & \textbf{Initialization Distribution} \\ \hline
\textit{N-RNN Parameters} & & \\
Base Recurrent Weights & $J_{xx}$ & $\mathcal{N}\left(0, \frac{1}{N_x}\right)$ \\
Input Projection Matrix & $B_{dx}$ & $\mathcal{N}\left(0, \frac{1}{N_u}\right)$ \\
Output Decoder Weights & $W_{qx}$ & $\mathcal{N}\left(0, \frac{1}{N_x}\right)$ \\
Initial Hidden State & $x^{(k)}_0$ & $\mathbf{0}$ \\ \hline
\textit{A-RNN Parameters} & & \\
Recurrent Weights & $W_{hh}$ & $\mathcal{N}(0, 0.001)$ \\
Observation Weights & $W_{oh}$ & $\mathcal{N}\left(0, \frac{1}{N_o}\right)$ \\
Recurrent Bias & $b_h$ & $\mathbf{0}$ \\
Initial Hidden State & $h_0$ & $\mathbf{0}$ \\ \hline
\textit{Modulatory \& Policy Heads} & & \\
Modulatory Projection & $H_{px}$ & $\mathcal{N}(0, 0.01^2)$ \\
Policy Layer 1 Weights & $W_1$ & $\mathcal{N}(0, 0.01^2)$ \\
Policy Layer 1 Bias & $b_1$ & $\mathbf{0}$ \\
Policy Layer 2 Weights & $W_2$ & $\mathcal{N}(0, 0.01^2)$ \\
Policy Layer 2 Bias & $b_2$ & $\mathbf{0}$ \\ \hline
\end{tabular}
\end{table}

\begin{table}[h!]
\centering
\caption{\textbf{Model training and architecture hyperparameters.}}
\label{tab:hyperparameters}
\begin{tabular}{lll}
\hline
\textbf{Parameter} & \textbf{Symbol} & \textbf{Value} \\ \hline
\textit{Task \& Environment} & & \\
Number of Latent Contexts & $M$ & $4$ \\
Number of Astrocytic Policies & $N$ & $4$ \\
Baseline Sub-optimal Reward Prob. & -- & $(1-p^\star)/(N-1)$ \\ \hline
\textit{Optimization parameter} & & \\
A-RNN Learning Rate & $\alpha_{\text{A-RNN}}$ & $3 \times 10^{-4}$ \\
N-RNN Learning Rate & $\alpha_{\text{N-RNN}}$ & $1 \times 10^{-3}$ \\\hline
\textit{RL hyperparameters (A2C+GAE)} & & \\
Discount Factor & $\gamma$ & $0.98$ \\
GAE Parameter & $\lambda$ & $0.95$ \\
Entropy Coeff. & $c_2$ & $0.01$ \\
Value Loss Coeff. & $c_1$ & $0.50$ \\
Curriculum Augmentation Phase & -- & $500$ trials ($r_k = 10$) \\
Curriculum Annealing Phase & -- & $250$ trials ($r_k \to 1$) \\ \hline
\end{tabular}
\end{table}

\subsection{Training Paradigm and Curriculum Schedule} \label{app:sched}

As introduced in the main text, we decouple the optimization without introducing credit-assignment instability. To do so, we rely on a mixed regime, where the N-RNN is trained via supervised learning, and the A-RNN relies on an RL procedure over a trial-based schedule.
Because trial outcomes under probabilistic context switching are inherently ambiguous, early contextual action gradients can suffer from high variance. To stabilize value function estimation and accelerate convergence, we implement a curriculum schedule \cite{muller2025improving}:
\begin{itemize}
        \item Phase 1: Reward Augmentation (Trials 1--500): Correct contextual choices $\pi_k = c_k$ receive an augmented reward magnitude ($r_k = 10$).
        \item Phase 2: Reward Annealing (Trials 501--750): The reward scale is stepped down linearly ($r_k \to 1$) to allow the value head to smoothly adapt to unit reward magnitudes.
        \item Phase 3: Standard Sparse Evaluation (Trials 751+): Augmented rewards are completely removed. The A-RNN operates exclusively on unaugmented, sparse trial feedback ($r_k \in \{0, 1\}$).
    \end{itemize}

The A-RNN's parameters $\theta$ updated as in \cite{mnih2016asynchronous}:
    \begin{equation}
    \mathcal{L}_{\text{A-RNN}}(\theta) = \mathcal{L}_{\text{policy}}(\theta) + c_1 \mathcal{L}_{\text{value}}(\theta) - c_2 \mathcal{H}(\eta_\theta)
    \end{equation}
where $\mathcal{L}_{\text{policy}}(\theta) = -\mathbb{E}_k [\log \pi_\theta(a_k | h_k) \hat{A}_k^{\text{GAE}}]$, $\mathcal{L}_{\text{value}}(\theta)$ is the MSE loss on state-value estimates, $\mathcal{H}(\eta_\theta)$ is an entropy regularization term encouraging policy exploration. $\hat{A}_k^{\text{GAE}}$ is the multi-step advantage estimator as in \cite{schulman2015high}.

\subsection{A-RNN Recurrent Mask and Sparse Topology} \label{app:mask}
To enforce spatial specificity and structural sparsity within the slow A-RNN, its recurrent weight matrix $W_{hh} \in \mathbb{R}^{N_h \times N_h}$ is constrained by a static binary mask $M_{hh} \in \{0, 1\}^{N_h \times N_h}$ constructed upon model initialization.

Units in the A-RNN are mapped onto a spatial grid, and permissible recurrent connections are constrained to local receptive fields defined by a nearest-neighbor lattice topology. To achieve a target sparsity $\sigma \in (0, 1]$, a fraction $\sigma$ of these topological connections is selected to form the set of permissible connections $\mathcal{E}_{\text{active}}$, yielding:
\begin{equation}
    (M_{hh})_{ij} = 
    \begin{cases} 
    1, & \text{if } (i,j) \in \mathcal{E}_{\text{active}} \\ 
    0, & \text{otherwise} 
    \end{cases}
\end{equation}

All entries in $W_{hh}$ are initialized from a narrow Gaussian distribution $\mathcal{N}(0, 0.001)$. During optimization, gradient updates are masked such that only active recurrent pathways ($(M_{hh})_{ij} = 1$) are updated, while unselected entries remain locked near zero throughout training. This static sparse topology remains fixed across all training and evaluation phases.

\end{document}